\documentstyle[preprint,prb,aps,amstex,amssymb]{revtex}
\begin{document}

\draft

\title
{\bf Interplay of D-wave Superconductivity and Antiferromagnetism in the
Cuprate Superconductors: Phase Separation and the Pseudogap Phase Diagram 
}

\author{
 W. P. Su }

\vspace{0.15in}

\address{
Department of Physics and Texas Center for Superconductivity,
University of Houston, Houston, Texas 77204}
\maketitle

\begin{abstract}

\vspace{0.15in}

{\it To understand the interplay of d-wave superconductivity and 
antiferromagnetism in the cuprates, we consider a two-dimensional
extended Hubbard model with nearest neighbor attractive interaction. Free
energy of the homogeneous (coexisting superconducting and antiferromagnetic)
state calculated a s a function of the band filling shows a region of
of phase separation. The phase separation
caused by the intersite attractive force leads to novel insights into salient
features of the pseudogap phase diagram. In particular, the upper crossover
curve can be identified with the  phase separation boundary.
At zero temperature, the boundary constitutes a critical point. The
inhomogeneity observed in the underdoped cuprates is a consequence of
incomplete phase separation. The disorder (inhomogeneity) brings about the
disparity between the high pseudogap temperature and the low 
bulk superconducting
transition temperature.
 }
\end{abstract}
\pacs{PACS  numbers: 74.20.-z, 74.72.-h, 74.25.Ha}

\vspace{0.15in}

\noindent {\bf I. Introduction}\newline

The phase diagram~\cite{tim,tall}
of the cuprates highlights the unusual physics of the 
cuprates, as such it constitutes a touchstone of high-$T_{c}$ theory. 
Particularly intriguing is the mystery surrounding the nature of the
pseudogap phase. How is that phase related to the unusual superconducting
phase with the characteristic inhomogeneities~\cite{el}?
How are all those related to
the antiferromagnetic correlation which seems to coexist~\cite{sin,nie} with 
superconducitivty?

To address those questions, we consider an effective model for
antiferromagnetism and d-wave superconductivity: the t-U-V-W model.
This model~\cite{mic1,mrr1}
 is a slight extension of the familiar t-U-V model, which is the
regular Hubbard model supplemented with an intersite attractive
density-density interaction. It is not surprising that the t-U-V model yields
both antiferrogmanetic (AF) and d-wave superconducting (DSC) states. That is
the way it is supposed to work. What is somewhat surprsing is that the 
attractive intersite interaction can lead to phase separation (PS) in addition
to DSC. This has actually been recognized~\cite{lh}  before the discovery of the
cuprates. A detailed analysis of how phase separation affects the AF and DSC
properties of the system~\cite{castro,xing,kres,mello} 
within the t-U-V-W model has not been carried out before. As will be shown
below, such an analysis offfers much insight into the nature of the entire
phase diagram.

\vspace{0.15in}

\noindent {\bf II. The t-U-V-W Model}

The t-U-V-W model is defined as \vspace{0.1in}

\begin{eqnarray}
H=-t\sum_{<ij>\sigma}[c_{i\sigma}^{\dagger}c_{j\sigma}+H.c.]+
U\sum_{i}(n_{i\uparrow}-{\frac{1}{2}})(n_{i\downarrow}-{\frac{1}{2}})+
V\sum_{<ij>}(n_{i}-1)(n_{j}-1)  \nonumber \\
+\sum_{<i,j>^{\prime}}W_{i,j}(n_{i}-1)(n_{j}-1)-\mu\sum_{i}n_{i},
\end{eqnarray}

\vspace{0.1in}

where $n_{i\sigma}=c_{i\sigma}^{\dagger}c_{i\sigma}$ is a density operator
of the conduction electrons, $n_{i}=n_{i\uparrow}+n_{i\downarrow}$, $<ij>$
is a nearest neighbor pair and $<ij>^{\prime}$ is any other pair. 
We measure energy in units of t and set t=1. Since we calculate only
the properties of homogeneous states, the extended repulsive terms enters
only through W, which 
is the summation of $W_{ij}$ over all extended pairs $<ij>^{\prime}$
divided by the total number of lattice sites. 

The t-U-V-W model can be solved in the mean-field approximation by 
linearizing the interaction terms. In linearizing the attractive V term,
one arrives at three linear terms. The first is  a density term, the second
is a pairing term and the third is an exchange term. As opposed to most
previous calculations, we retain all three terms. For the W term, only the
density term is kept after linearization.
 As usual, self-consistent solutions are obtained by
iterations.
Details of the method and expression of the free energy are given in 
Micnas et al.'s paper~\cite{mrr}.\newline

\noindent {\bf III. Mean Field Results}

Following the general strategy of treating  phase separation in a binary
mixture~\cite{put}, we first evaluate the free energy of the homogeneous state as 
a function of average density (the filling fraction) $n$. For this calculation
we adopt the following parameter values $U=2.1$, $V=-0.9$~\cite{uv} and $W=0.4$~\cite{wterm}.
The squares in Figure 1 denote the free energy of the pure DSC state, the
circles that of the coexisting DSC and commensurate AF state at temperature
T=0. It is clear that for band filling larger than 0.42 DSC and AF coexist~\cite{kyun},
whereas for density less than that AF does not exist. 
A double-tangent construction yields the region between A and B as
the two-phase region. A and B represent the two terminal phases. In a 
regular binary system, the terminal phases A and B are the controlling
phases, the lowest energy state with a density between A and B consists of
an A-rich region separated from a B-rich region. This is a global phase
separation. In the cuprates, due to the charges of the random dopants, we can
expect only limited incomplete phase separation. STM results support
a picture of nanoscale phase separation, with a density variation over regions
of nanoscale~\cite{el}.

To examine the nature of the coexistence, we plot the AF and DSC order
parameters at zero temperarure
as a function of band filling in Figure 2. It is clear that
as the density approaches half filling, AF builds up rapidly but DSC continues
to stay strong until it is eventually suppressed by AF at half-filling.
Except for half filling, it seems as if AF is transparent to DSC. 
That is 
consistent with the muon spin relaxation experiment~\cite{sin}.
 The micro phase separation leads
to inhomogeneous disordered state in the phase separation region. Due to the
disorder, one expects a disparity between the global and local DSC 
correlations. 
In addition, the disorder
will disrupt the long-ranged AF order except for a narrow region close to
half filling. Due to density fluctuations, even for band fillings less than
0.42 AF correlation persists. As a matter of fact, the region of short-range
AF correlation can extend all the way to the crtitical point A, the
PS boundary point~\cite{pana}. 
Beyond that only pure homogeneous DSC state is possible.
In our viewpoint A corresponds to the quantum critical point~\cite{tall,pana}
 discussed in
the literature.

The calculation of free energy can be repeated for finite temperatures. From
those calculations, we compile a phase diagram shown in Figure 3. The squares
indicate the transition temperature of the DSC state in the absence of phase
separation. The circles and triangles delineate the phase separation boundary.
The regions of DSC outside of the PS boundary are immune to PS, they are 
the regular homogeneous DSC. Inside the PS boundary 
the bulk $T_{c}$ will be modified as in a granular
superconductor~\cite{kres,dynes}
 leading to the disparity between the local DSC ordering
temperature (the square) and the actual $T_{c}$. Thus our result supports
the precursor pairing picture~\cite{carlson}
of the pseudogap phase of Kivelson et al.
The density fluctuations associated with PS can spread the effect of AF
correlation over the entire PS region. This can explain the widespread
presence of AF correlation seen in the muon spin relaxation experiment~\cite{pana}.

A comparison of Figure 3 with the phase diagrams~\cite{tim,tall}
 proposed so far naturally
identifies the phase separation boundary with the upper crossover curve for the
hole-doped cuprates. The transition temperature curve of the 
metastable homogeneous DSC in Figure 3 can then be regarded
as the lower crossover curve. The latter identification is not as clearcut as the
former because the lower crossover curve does not even exist in certain versions
of the phase diagram. Nonetheless, we regard it as a physical crossover into a region
(the pseudogap region) where at least the local electronic~\cite{ding}
 and magnetic properties~\cite{panap}
of the system resemble those of a superconductor. This crossover usually happens at a 
significantly lower temperature than the upper crossover temperature.

The model Hamiltonian (1) considered so far has the electron-hole conjugation
symmetry  which we know does not exist in real cuprates. To implement this
asymmetry, we include a next-nearest neighbor hopping 
term (with amplitude $t'$) in the Hamiltonian. We
recalculated the order parameters at zero temperature. The result is displayed
in Figure 4 for $t'=-0.15t$. As before, the squares and circles denote DSC 
and AF order parameter respectively. A and B mark the phase separation
boundary points. Compared to Figure 2, the terminal phases A and B have 
 been shifted to lower band fillings.  
In the electron-doped case the
DSC is significantly weakened compared to the hole-doped case. 
There is  very little room (small density range) for
the pseudogap phenomenology to develop. 
Thus a simple extension of the hopping term
in the Hamiltonian seems to be able
to capture the contrast between the phase digrams of the hole-doped
and the electron-doped cuprates as seen experimentally~\cite{shen}.
Since the pseudogap phenomenon is much better characterized in the hole-doped 
samples than in the electron-doped samples, we will focus on the former in the
following.

While much remains to be pursued, it is clear that even at the mean-field
level, the t-U-V-W model is very promising as a phenomenological model
of the cuprate phase diagram. That suggests that an explanation of the
phase diagram requires only the existence of an intersite pairing force
independent of the microscopic origin of the pairing.\newline

\vspace{0.15in}

\noindent {\bf IV. Quantum Monte Carlo Simulation of the t-U-V Model} 

For a limited region of the parameter space, we have carried out some
quantum Monte Carlo calculation~\cite{wpp}
of the t-U-V model (W=0) to validate the mean
field result. Phase separation is clearly demonstrated by a break in the
density versus chemical potential curve. Due to finite size of the 
simulated system, a complete global phase separation in real space is
not possible. Nonetheless, the density fluctuations are clearly seen in
the density histograms (Fig 5 of Ref. 24).

\vspace{0.15in}

\noindent {\bf V. Excitonic Origin of Intersite Pairing: the Little Model}

Soon after  Bardeen, Cooper, and Schrieffer~\cite{BCS}
 (BCS) proposed their theory of  phonon-based
superconductivity, the search has been underway for an alternate mechanism for
superconductivity. In 1964, Little proposed a specific model of electronic (excitonic)
origin.~\cite{little}. He considered a quasi-one-dimensional conductor with densely
packed polarizable molecules surrounding the conducting spine as a most likely candidate
of excitonic superconductors.

We have adopted a tight-binding version~\cite{HS} of the Little model. For an
onsite electron-exciton (repulsive) coupling, quantum Monte Carlo simulations
suggest~\cite{su3,min} that in two dimensions, s-wave superconductivity can exist
except near half filling where charge-density-wave dominates. As such the Little
model is a viable model of electronic mechanism of superconductivity.

By extending the range of electron-exciton coupling, one would expect effective
intersite attractive interactions between the conduction electrons induced by
electronic polarization. Preliminary quantum Monte Carlo data indeed shows signs
of d-wave pairing and phase separation.
 
\vspace{0.15in}

\noindent {\bf VI. Discussion}

High-temperature superconductivity has been one of the most vexing problems in
condensed matter physics. Most of the theoretical effort up to date starts with the
insulating AF state and seeks to `derive' the superconducting properties of the
doped states. This is a huge attempt which has so far met with little success.

We have adopted a less ambitious
phenomenological approach. We have  assumed the existence of
an intersite attractive  plus onsite as well as extended  repulsive potential
(as embodied in the t-U-V-W model), and then tried to work out the consequences.
It turns out that even at the mean field level, the model yields phenomenologies
strikingly similar to what have been observed in the cuprates. In particular, we
see that an intersite attraction leads to not only d-wave superconductivity, but
also phase separation. The PS strongly modifies the coexistence of AF and DSC.
Density fluctuation and disorder associated with PS profoundly affect the nature of
the states within the phase separation boundary. While more details remain
to be worked out to prove that this is `the' explanation of the phase diagram
and the pseugogap state, it is clear that the effect of PS has to be taken seriously
along with the existence of DSC, as they share the same origin.

Our theory is simple outside of the PS boundary, it gets more complicated as one moves
deep into the PS region where SDW mixes with DSC. It would indeed be  
a daunting task to go all the way from the pure AF state
to reach the simple homogeneous DSC state without assuming an intersite pairing force.

In our view, the phase diagram can be understood in terms of the t-U-V-W model, thus
leaving the origin of the pairing force open. That is a more subtle question
as it might depend on  the material parameters. It is 
probably going to take much more theoretical effort to nail that down. Nonetheless we have
provided  a possible microscopic model of pairing and PS. It is a tight-binding version of
the good old Little model. Even if it turns to be irrelevant for the cuprates, it could
still serve as a nontrivial and
important model of the electronic mechanism of superconductivity. As such it deserves further
study.

This work was partially supported by the Texas Center for Superconductivity,
 the Robert A. Welch Foundation (grant number E-1070),
and the National Science Council of Taiwan under contract number NSC
93-2112-M-110-011. Long term
support of C. W. Chu and C. S. Ting is gratefully acknowledged. Thanks are
also due to I. M. Jiang and the Department of Physics at the National
Sun Yat-Sen University for their hospitality durng the summer of 2005.

\begin{figure} 

\noindent Figure 1.
Free energy (per site)
of the homogeneous d-wave superconducting state (squares)
and that of the homogeneous coexisting SDW and DSC state (circles) at zero temperature in the
t-U-V-W model.
\vspace{0.15in}

\noindent Figure 2.
Spin-density-wave order parameter (spin-up density minus spin-down density) and D-wave 
superconducting 
gap order parameter $\Delta_{d}=|\frac{V}{2}<c_{i\uparrow}c_{j\downarrow}
-c_{i\downarrow}c_{j\uparrow}>|$ 
(i,j is a nearest
neighbor pair) as a function of band filling in the t-U-V-W model.
\vspace{0.15in}

\noindent Figure 3.
Illustration of the interplay between superconductivity and phase
separation. The squares are superconducting transition temperatures in the
absence of phase separation.
\vspace{0.15in}

\noindent Figure 4. Same as in Figure 2, with a next-nearest neighbor hopping term added to
the Hamiltonian (1).
\vspace{0.15in}

\end{figure}



\vspace{0.3in}


\begin{thebibliography}{99}
\bibitem{wterm} 
The $W$ term contributes  an amount 
$4W(n-0.5)^{2}$ to the free energy.
For $W=0$ the  minima of the free energy would be
lowered to -1.533 (for $n=0.2$ and $n=0.8$). The antiferromagnetic state 
at $n=0.5$ would no longer be a stable
terminal phase.
\end{thebibliography}
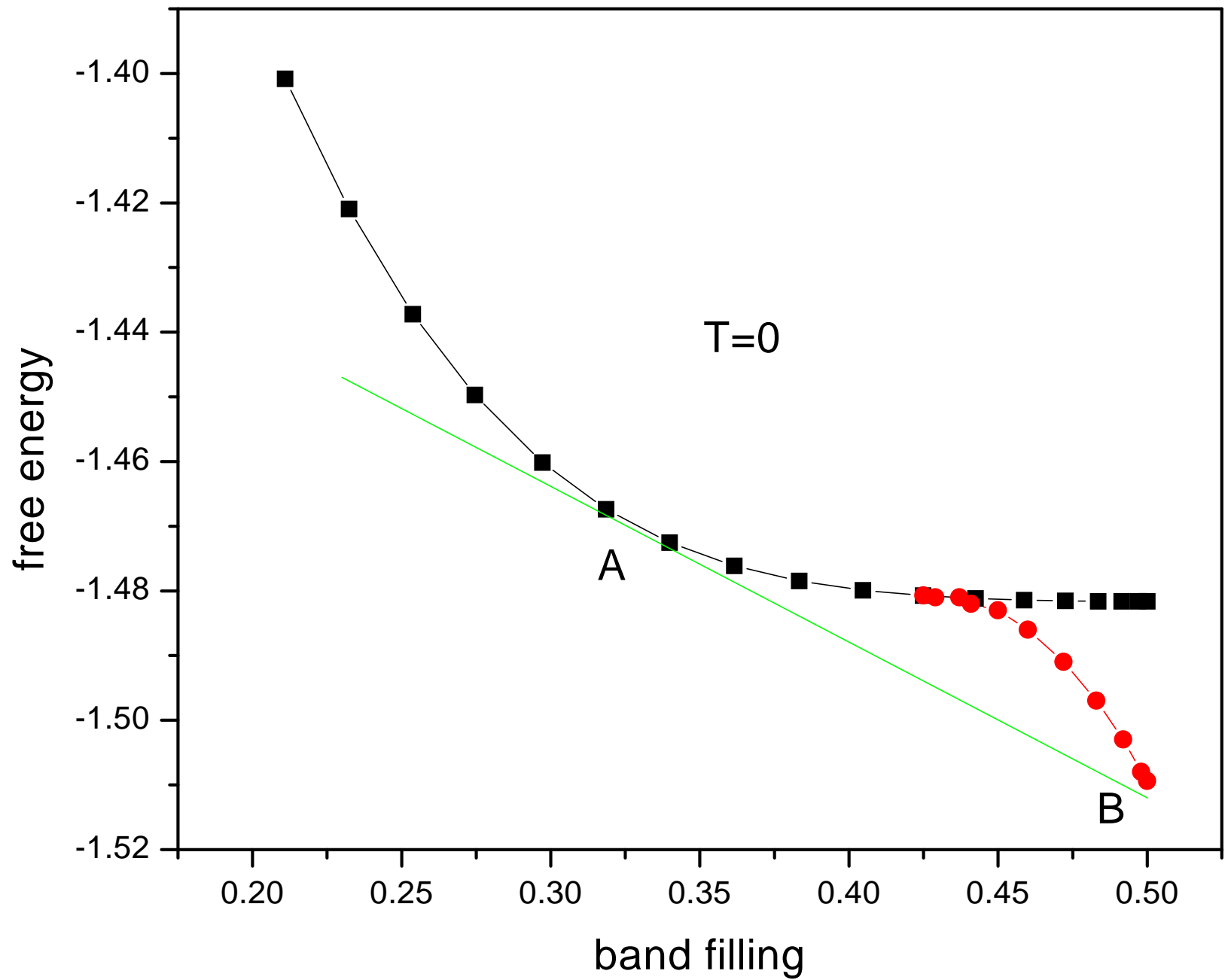
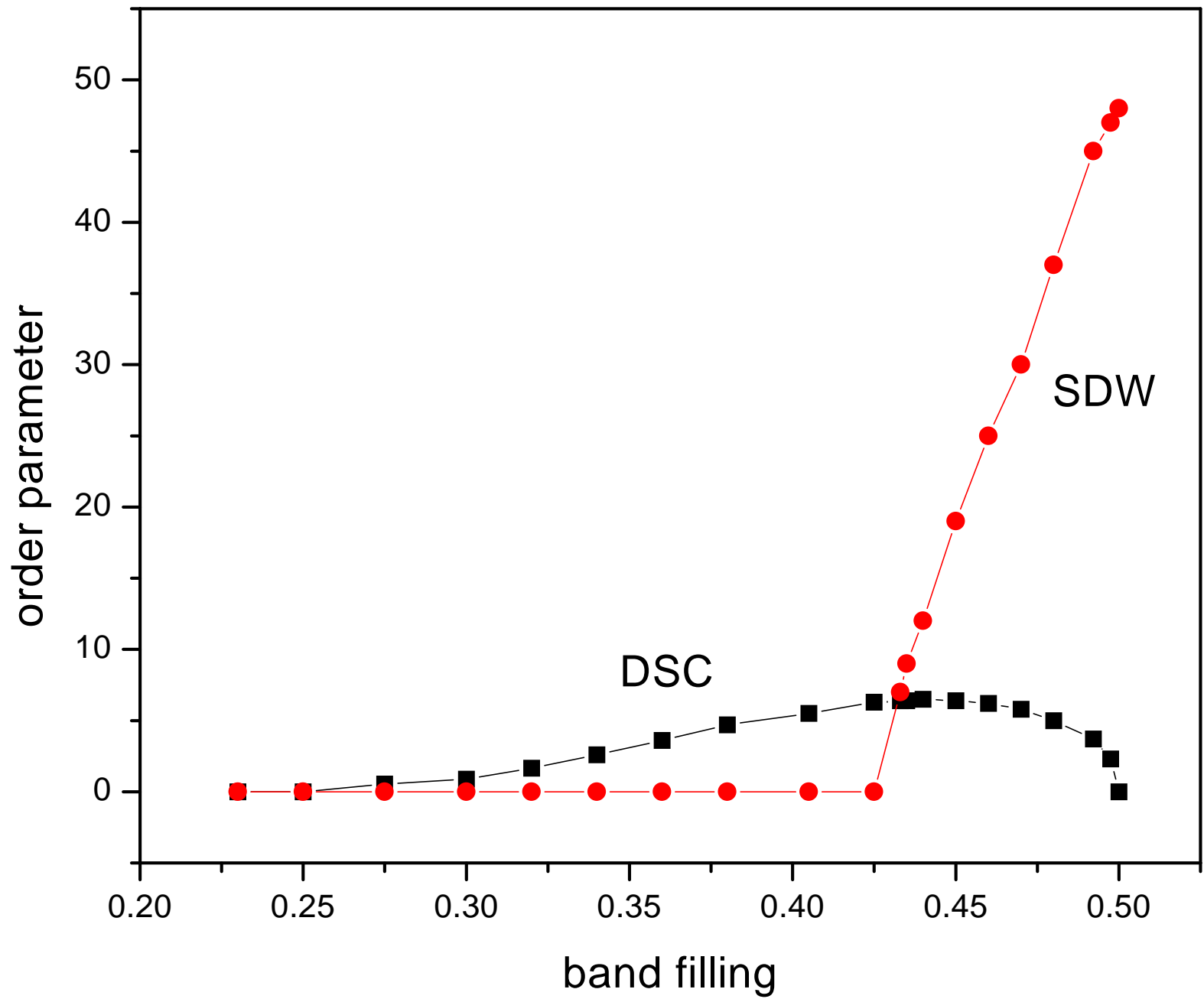
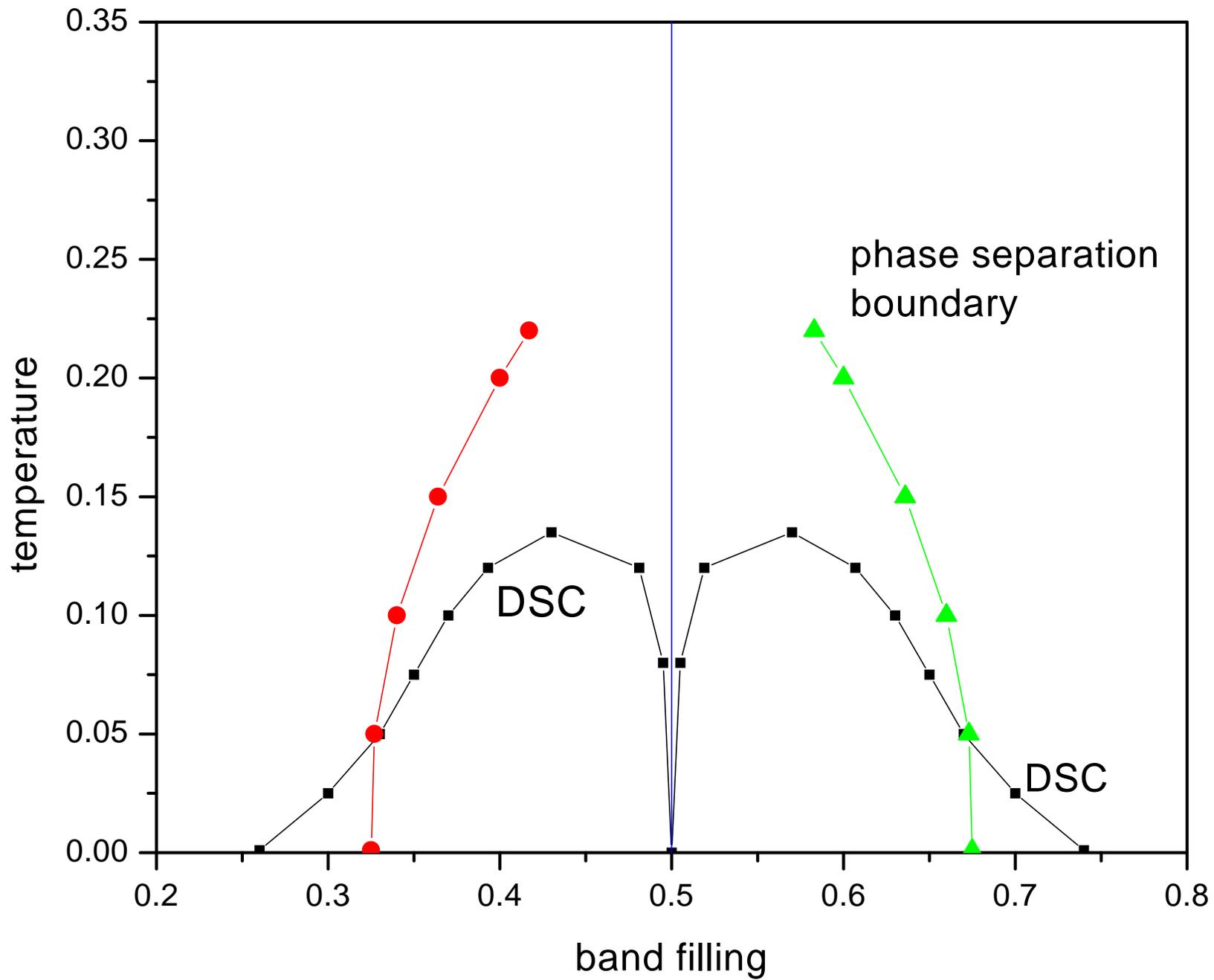
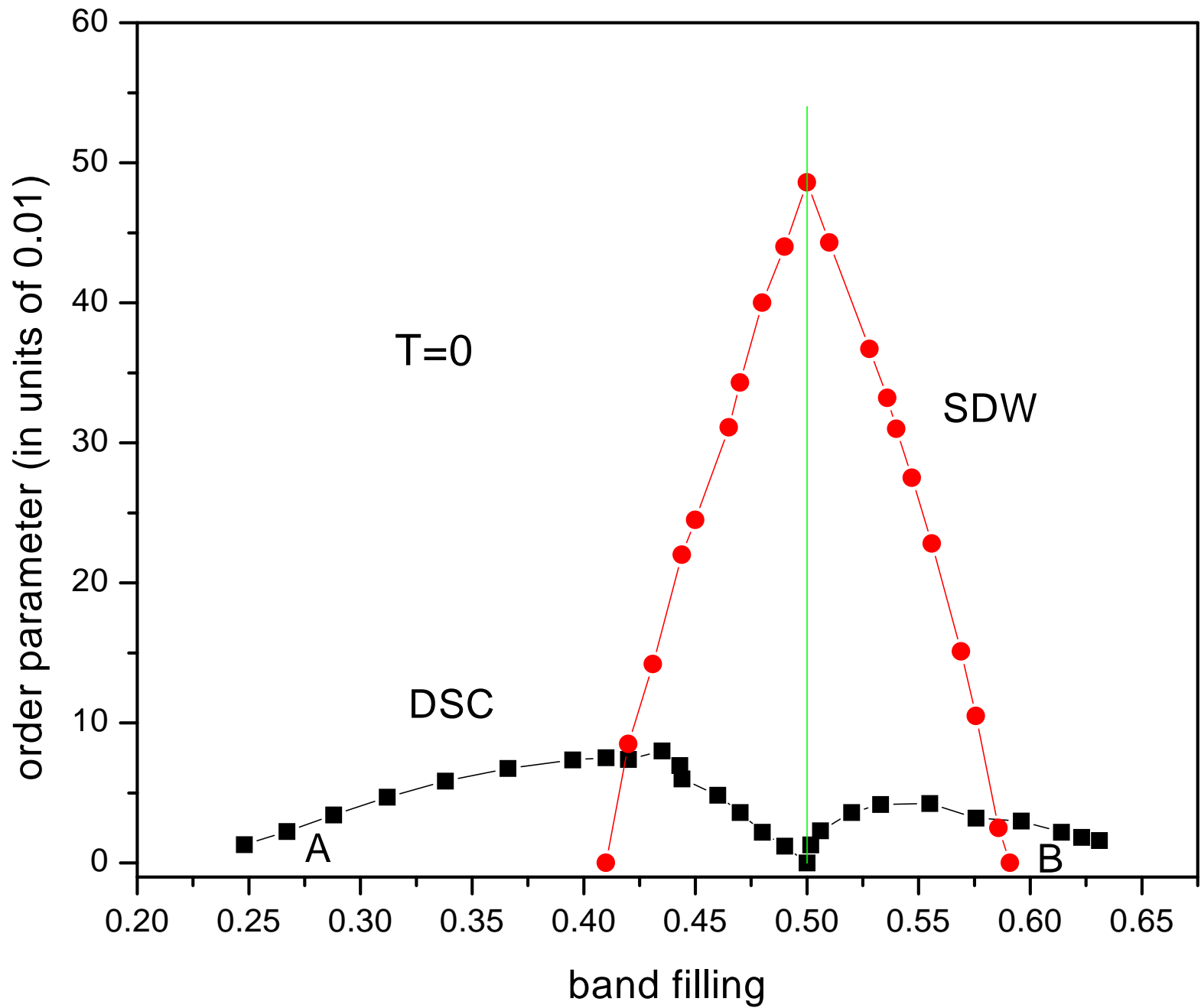
\end{document}